\documentclass[%
 reprint,
 amsmath,amssymb,
prstab,
]{revtex4-2}
\usepackage{color,soul}
\usepackage{graphicx}
\usepackage{dcolumn}
\usepackage{bm}
\usepackage{natbib}
\usepackage{mathtools, nccmath}
\begin{document}

\newcommand{\beginsupplement}{%
        \setcounter{table}{0}
        \renewcommand{\thetable}{S\arabic{table}}%
        \setcounter{figure}{0}
        \renewcommand{\thefigure}{S\arabic{figure}}%
     }

\preprint{}

\title{Failure of Conventional Roughness Metrics in Assessing Field-Limiting Mesoscopic Topography in SRF Nb Films on Cu}


\author{Eric M. Lechner\,$^{1*}$, Brandi Redman\,$^{2}$, Theodore S. Cahall\,$^{3}$ and  Anne-Marie Valente-Feliciano\,$^{1}$}
  \affiliation{Thomas Jefferson National Accelerator Facility, Newport News, Virginia 23606}
  \affiliation{Department of Physics, Old Dominion University, Norfolk, Virginia 23529}
  \affiliation{Department of Physics, Oregon State University, Corvallis, Oregon 97331}

\thanks{lechner@jlab.org}%

\date{\today}

\begin{abstract}

A characteristic corrugated surface morphology of Nb films on Cu is identified, with mesoscopic features comparable to the London penetration depth and coherence length, and its impact on superconducting radio-frequency cavity performance metrics is examined. Magnetic field enhancement factors and superheating field suppression factors are calculated within the London model for a representative corrugated geometry. These results demonstrate that roughness trends derived from buffered chemical polished and electropolished Nb cavities do not capture the impact of nanoscale surface morphology on SRF Nb thin film performance, as even surfaces that possess low average roughness can contain geometric features that generate substantial local field enhancement and significantly suppress the Bean–Livingston barrier. The influence of surface roughness on impurity diffusion is also investigated, which highlights the roles of increased surface area and local geometric confinement in modifying near-surface impurity distributions. Tracking the effect of impurity diffusion on the evolution of magnetic field enhancement, we show that geometrically confined impurity distributions can mitigate nanoscale magnetic field enhancement substantially.

\end{abstract}

\maketitle


\section{\label{sec:level1} Introduction}
Superconducting radio frequency (SRF) cavities play a central role in high duty cycle and continuous wave particle accelerators due to their ability to sustain strong electromagnetic fields with minimal energy dissipation \cite{reece2016continuous,SingerReview2016,Dhakal2024SNSPPUPerformance,Maniscalco20231LCLSIIHE,Wei2019TheFRIBSCLinacInstallAndComissioning}. These cavities are often fabricated from bulk Nb or Nb-coated Cu. Nb deposited on Cu is of significant interest because Cu is relatively cheap compared with SRF cavity-grade Nb while offering superior machinability and thermal conductivity. In SRF applications, only the top few hundred nanometers of the surface participate in RF screening, so a thin high‑quality Nb layer, typically 1-10 $\mu$m thick, is sufficient \cite{valente2016superconducting}. This makes thin film structures an effective route to reducing SRF accelerator capital costs relative to high‑purity bulk Nb while still delivering high performance. Realizing high performance Nb/Cu cavities requires control of film morphology and interface properties, as surface defects, stress, and substrate induced roughness can severely limit high-field performance \cite{Hryhorenko2026ElectropolishingInducedDefect,Xie2011QuenchSimulation}. As such, understanding and improving the topography and microstructure of Nb thin films on Cu is essential for their successful adoption in next generation accelerator technologies \cite{Hryhorenko2026ElectropolishingInducedDefect}. 

The performance of SRF cavities is tied to the condition of their inner surfaces. Among other factors, surface roughness, even at the nanometer scale, can weaken the metastability of the Meissner state, leading to low-field vortex penetration and rapidly increasing RF dissipation. These effects limit the attainable accelerating gradient and high-field quality factor ($Q_0$) of SRF cavities \cite{gurevich2008dynamics,xu2016simulation}. The surface processing history of Nb cavities dramatically affects their SRF performance. In particular, buffered chemical polishing (BCP) and electropolishing (EP) produce characteristically different surface morphologies. Such features are understood to enhance the local magnetic field, suppress the superheating field ($B_{\text{sh}}$), and promote low-field thermal instability. The same effects have also been identified in Nb$_3$Sn, where thermally grooved grain boundaries formed during vapor diffusion \cite{Lechner2025Nb3SnTopo}. At the level of average surface roughness, bulk Nb appears to follow an intuitive trend: BCP-treated surfaces are rougher ($R_a \approx$ few $\mu$m \cite{tian2006surface,Hryhorenko2026ImpactOfMetallographicPolishingOnRFPerformance}) and generally show poorer high-field performance, whereas EP-treated surfaces are smoother ($R_a \approx100$ nm \cite{prudnikava2018}, with the height of grain boundary defects introduced by the electropolishing process $\approx 50-60$ nm \cite{Hryhorenko2026ElectropolishingInducedDefect}) and reach substantially higher fields. Nb films on Cu, however, do not fit neatly into this picture. Their average surface roughness is low ($R_a$ is in the tens of nm range \cite{Ries2020SuperconductingPropertiesAndSurfaceRoughnessOfThinFilmNb,Ries2023MagneticFieldEnhancementandImpactofSurfaceDefects,Turner2023SuperconductingPropertiesSurfaceMorphologyofSputteredNbFilmsonCu}), yet the most exceptional Nb/Cu cavities have only reached peak magnetic fields as high as 120 mT \cite{VenturiniDelsolaroSeamlessQWRForHIEISOLD}. Given these observations, can the effects of surface roughness be ruled out as the leading factor in high-field limitations of Nb films on Cu?  

Motivated by this question, we investigate the effect of nanoscale topography common in Nb deposited on Cu films. Using atomic force microscopy (AFM), we identify a characteristic corrugated surface morphology and quantify the effects of this geometry on magnetic field enhancement and superheating field suppression within the London model. Since the London model describes a rigid Meissner state, the calculations made in this work provides a conservative estimate of roughness-induced field enhancement and superheating field suppression. Together, these results show that conventional roughness metrics are insufficient to account for the SRF performance of Nb/Cu films when compared with BCP and EPed bulk Nb surfaces. 
\section{\label{sec:level1} Methods}
\subsection{Atomic Force Microscopy}
Atomic force microscopy measurements were made in tapping mode using a Digital Instruments Dimension 3100 atomic force microscope equipped with a Nanoscope IV controller on loan to Thomas Jefferson National Accelerator Facility from The College of William and Mary. The silicon AFM probe features a tip radius less than 10 nm. Topographic images were acquired in tapping mode (TM) over an area of 10 $\mu$m $\times$ 10 $\mu$m, 5 $\mu$m $\times$ 5 $\mu$m, and 1 $\mu$m $\times$ 1 $\mu$m consisting of 512 $\times$ 512 pixels. The tip hosts a half-cone angle between 20° and 25° along the cantilever axis, 25° and 30° from the side and 10° at the apex.
\subsection{Sample Preparation}
The Nb film examined in this work was prepared at Jefferson Lab on an OFHC Cu substrate by electron cyclotron resonance (ECR) plasma energetic condensation. Before deposition, the substrate was mechanically polished, electropolished, and heated at 360 °C for 24 h to reduce or dissolve the native Cu oxide and promote the formation of a crystalline Nb/Cu interface. The heteroepitaxial Nb film was then grown at 360 °C, beginning with a 5 min nucleation step at an ion energy of 244 eV, followed by 37 min of subsequent growth at 64 eV.

\section{\label{sec:level1} Results}
\subsection{Topographic Characterization of Nb films}
\begin{figure}[!h]
    \centering
    \includegraphics[width=1\linewidth]{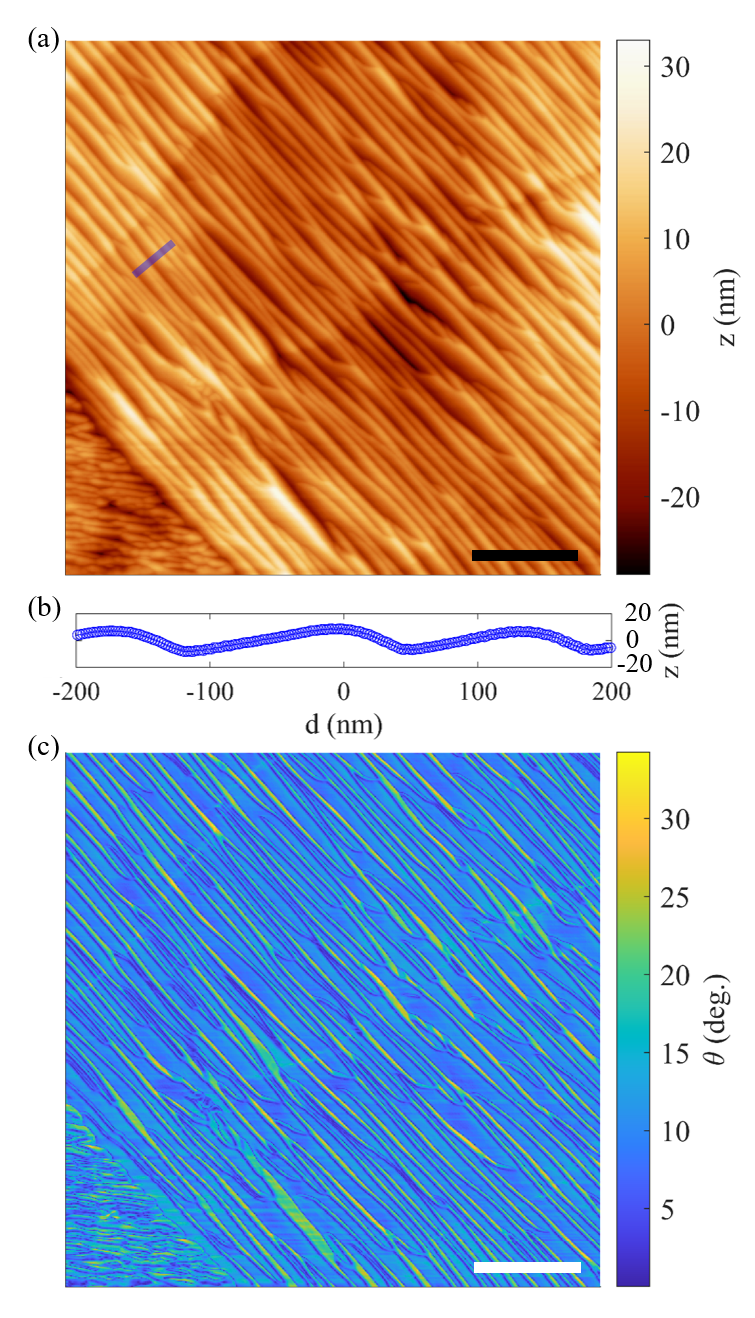}
    \caption{(a) Representative TM-AFM topography, (b) line profile derived from the blue area shown in (a), and (c) local slope angle of the corrugation geometric defect. The scale bar in the images are 1 $\mu$m.}
    \label{fig1}
\end{figure}

Tapping mode atomic force microscopy (TM-AFM) was utilized to examine the surface roughness of the Nb films deposited on Cu. A representative TM-AFM topograph over a 5 $\mu$m $\times$ 5 $\mu$m area is shown in Fig. \ref{fig1} (a). For this surface, $R_a =$ 6.7 nm and $R_q =$ 8.3 nm. The surface is characterized by corrugations which roughly approximate a triangle wave geometry. These corrugations have been observed in a number other studies \cite{Catani2007DepositionCharacterisationofNiobiumFilms,Valizadeh2021SynthesisofNbandAlternativeSuperconductingFilmtoNbforSRFCavityasSingleLayer,Turner2023SuperconductingPropertiesSurfaceMorphologyofSputteredNbFilmsonCu}. The peak-to-valley variations, which define the color scale in Fig. \ref{fig1} (a), are on the scale of a few tens of nanometers. A line profile constructed from a thin area perpendicular to the ridges indicated by the blue line in Fig. \ref{fig1} (a), shown in Fig. \ref{fig1}(b), highlights the scale of the waves. The corrugations have a wavelength of $\approx$150 nm and an amplitude of $\approx$30 nm. These dimensions, which are comparable to the London penetration depth ($\lambda$ $\approx$ 30 nm and $\xi$ $\approx$ 40 nm in clean-limit Nb \cite{McFadden2026NbCoherenceLengthAndPenetrationDepth}), and the sharp V-shaped groove and are sufficient to induce significant current crowding, magnetic field enhancement, and degradation of the Bean-Livingston barrier. 

The local slope angle map calculated from the topography as described in Ref. \cite{Lechner23TopographicEvolution}, is presented in Fig. \ref{fig1}(c). Fig. \ref{fig1} (c) presents a mosaic-like map of slope angles reaching above 30°. These angles are comparable in magnitude to the high slope angles also observed at intergranular steps on electropolished bulk niobium which can account for a primary field-limitation in SRF cavities \cite{Hryhorenko2026ElectropolishingInducedDefect}. However, because of the corrugated topography, the internal angle at the corrugation trough is larger than that encountered in the electropolished case. In addition, these corrugations modify shallow impurity diffusion which can be important for high-field performance.

\subsection{Comments on Some Surface Roughness Metrics}

Average roughness, $R_a$, is a common surface roughness metric used in many disciplines. It is defined by the deviation of the absolute z displacement of the surface from its mean plane. In the case of a triangle wave geometry,  $z=\delta(1-4|x\tan(\theta)/2\delta -\lfloor x\tan(\theta)/2\delta +1/2\rfloor|)/2$, where $\delta$ is the corrugation height, the surface roughness is invariant on $\theta$,

\begin{equation}
R_a = \frac{\tan(\theta)}{2\delta} \int_0^{\frac{2\delta}{\tan\theta}} |z(x,\theta)| \, dx = \frac{\delta}{4}.
\label{CommentOnUseOfRa}
\end{equation}
Since superheating field suppression and magnetic field enhancement depends on more than just the height variation across the surface \cite{kubo2015field}, the invariance of average roughness to the slope angle renders average roughness an ineffective metric for properly accounting for the effect of surface roughness on field limitations. A similar expression can be obtained for RMS roughness ($R_q$)
\begin{equation}
R_q = \sqrt{\frac{\tan(\theta)}{2\delta}\int_0^{\frac{2\delta}{\tan\theta}} z^2(x,\theta) \, dx}  = \frac{\delta}{\sqrt{12}}.
\label{CommentOnUseOfRq}
\end{equation}
as well as $R_v$, $R_p$ and $R_z$ defined by
\begin{equation}
R_v = \min(z(x,\theta))  = -\delta/2,
\label{CommentOnUseOfRv}
\end{equation}
\begin{equation}
R_p = \max(z(x,\theta))  = \delta/2,
\label{CommentOnUseOfRp}
\end{equation}
\begin{equation}
R_z = R_p-R_v  = \delta,
\label{CommentOnUseOfRz}
\end{equation}
Finally, skew and kurtosis, which quantify whether the height distribution is biased toward peaks or valleys and how heavily-tailed the height distribution is, respectively,
\begin{equation}
R_{sk} = \frac{1}{R_q^3}\frac{\tan(\theta)}{2\delta} \int_0^{\frac{2\delta}{\tan\theta}} z^3(x,\theta) \, dx = 0,
\label{CommentOnUseOfRsk}
\end{equation}
\begin{equation}
R_{ku} = \frac{1}{R_q^4}\frac{\tan(\theta)}{2\delta} \int_0^{\frac{2\delta}{\tan\theta}} z^4(x,\theta) \, dx = \frac{9}{5},
\label{CommentOnUseOfRku}
\end{equation}
also contain no dependence on slope. These quantities describe the distribution of surface heights but not their lateral arrangement. They therefore cannot distinguish broad, shallow corrugations from steep, closely spaced corrugations with the same peak-to-valley depth. Even metrics that contain some slope information, such as RMS slope, surface area, or power spectral density, do not contain the necessary structure required to connect surface morphology to magnetic field enhancement, Bean-Livingston barrier suppression, and impurity redistribution.

\subsection{\label{subsec:SFS} Superheating Field Suppression at Corrugations}

\begin{figure}[!h]
    \centering
    \includegraphics[width=1\linewidth]{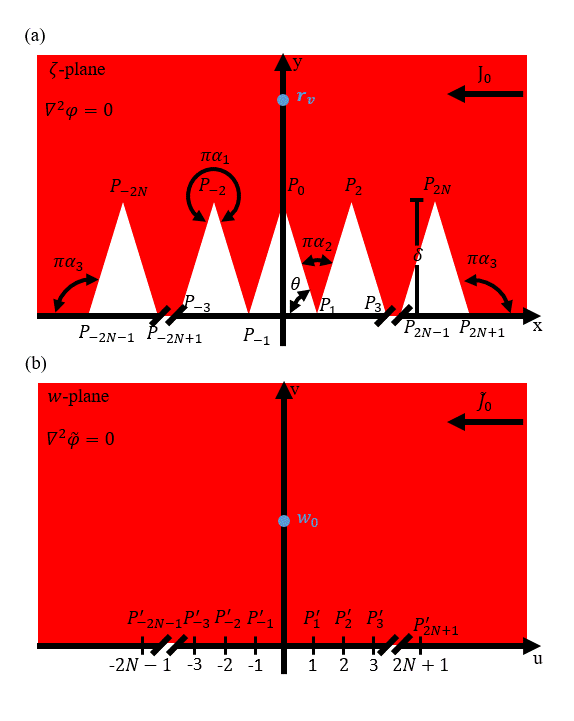}
    \caption{(a) Real-space representation of the repeated corrugation geometry and (b) the transformed geometry on the complex plane.}
    \label{CorrugatedConformalMappingGeometry}
\end{figure}

Surface roughness reduces the strength of the Bean-Livingston barrier. To estimate the effect of surface roughness on the superheating field we consider a superconductor in the extreme type-II limit and consider the balance of forces on a penetrating vortex. The force pushing the vortex into the superconductor arises from the external magnetic field is $\mathbf{F_M}$ and the force from the surface which expels it, $\mathbf{F_S}$ which arises due to the boundary condition of $\mathbf{J\cdot n=0}$. Here $\mathbf{F_M}=\mathbf{J_M} \times \phi_0\mathbf{\hat{z}}$ and $\mathbf{F_S}=\mathbf{J_I} \times \phi_0\mathbf{\hat{z}}$, where $\mathbf{J_M}$ is the screening current density and $\mathbf{J_I}$ is the current density of an image vortex which provides the force attracting the penetrating vortex towards the surface and accounts for the boundary condition at the surface. The Bean-Livingston barrier becomes unstable to vortex penetration when the sum of the forces on the vortex vanishes, $\mathbf{F_S} + \mathbf{F_M}=\mathbf{0}$, \cite{kubo2015field}. In the extreme type-II limit, $\nabla \cdot \mathbf{J} = 0$ and $\nabla \times \mathbf{J} \approx 0$ which allows conformal mapping to be employed to determine the complex potential for the screening current density ($\mathbf{J_M}=-\nabla \Phi_M$) at the geometrical defect as well as the complex potential for the current density of the image vortex ($\mathbf{J_I}=-\nabla \Phi_I$)\cite{buzdin1998ElectromagneticPinningOfVortOnDefects, aladyshkin2001best,kubo2015field}. To model our surface, we define a triangle-wave geometry where the slope angle $\theta$ is defined from the x-axis as shown in Fig. \ref{CorrugatedConformalMappingGeometry}. The triangle-wave geometry requires that three angles be defined for the conformal mapping calculation. In this case, the three interior angles required for the calculation are determined from $\theta$ by the following relations.

\begin{equation}
\pi\alpha_1=\pi+2\theta
\label{InteriorAngleAlpha1}
\end{equation}
\begin{equation}
\pi\alpha_2=\pi-2\theta
\label{InteriorAngleAlpha2}
\end{equation}
\begin{equation}
\pi\alpha_3=\pi-\theta
\label{InteriorAngleAlpha3}
\end{equation}

\begin{figure*}
    \centering
    \includegraphics[width=1.0\textwidth]{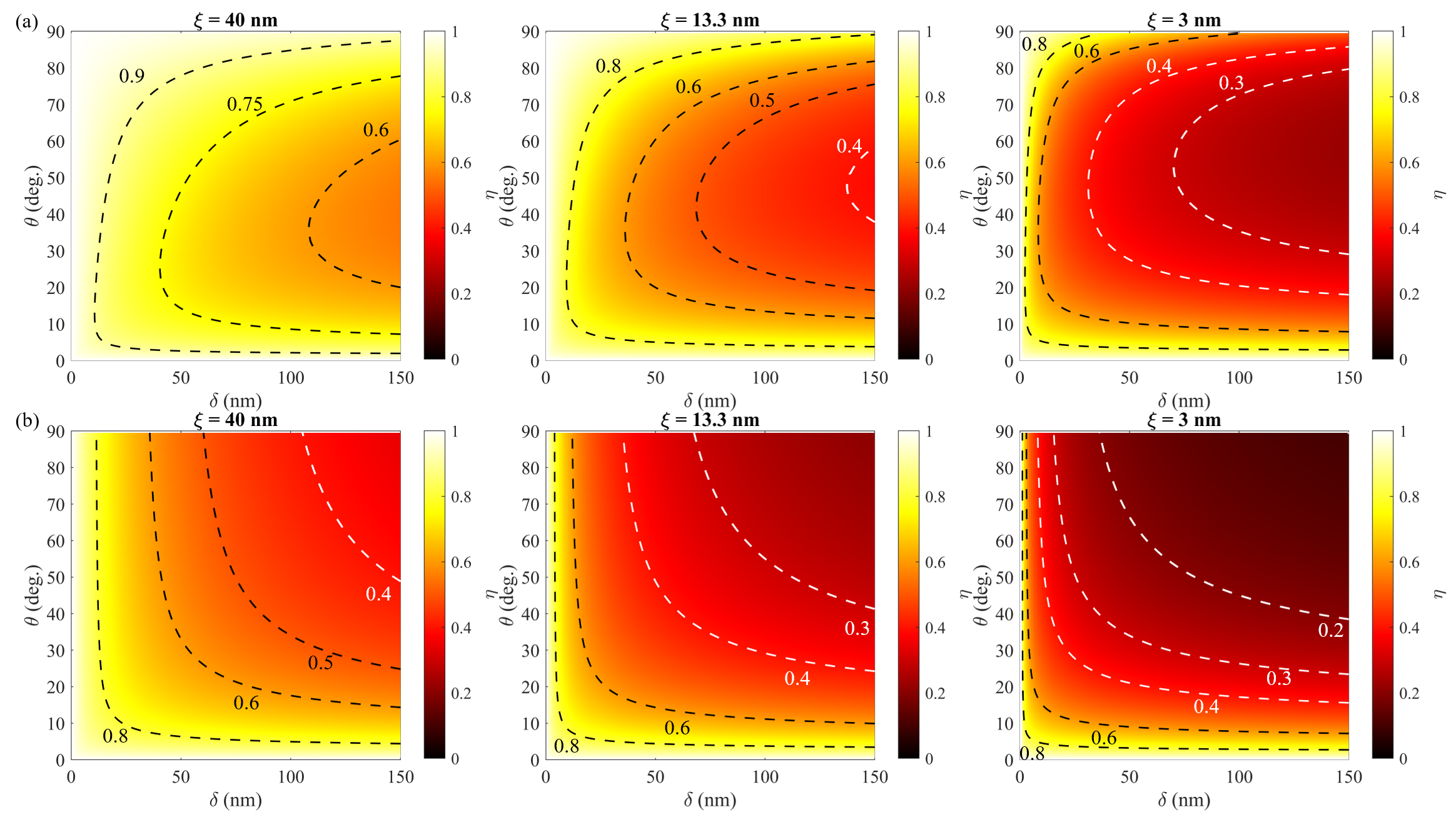}
    \caption{(a) Effect of corrugated surface roughness on superheating field suppression factor for $\xi = 40, 13.3$ and $3$ nm. (b) Effect of isolated grooves on superheating field suppression factor for $\xi = 40, 13.3$ and $3$ nm.}
    \label{EffectOfGeometryOnSFS}
\end{figure*}

\begin{figure}[!h]
    \centering
    \includegraphics[width=1\linewidth]{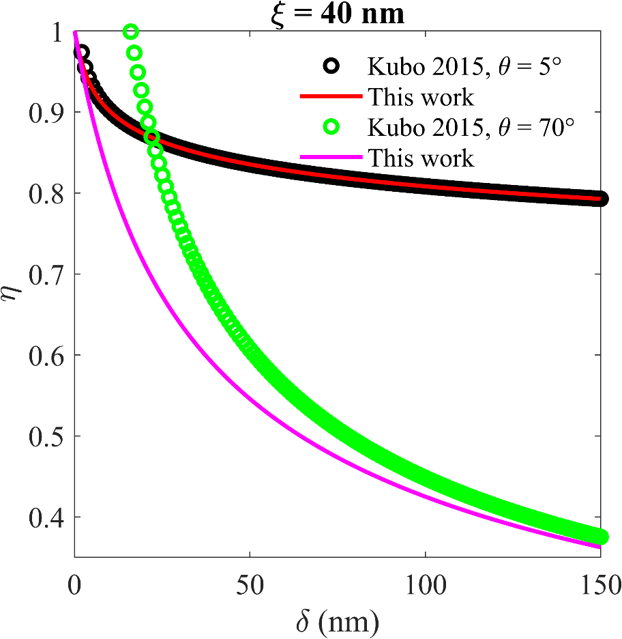}
    \caption{Effect of single groove geometry on the SFS factor within the limits described in Ref. \cite{kubo2015field} (black circles) and outside those limits at high slope angle (green circles).}
    \label{ComparisonWithKubo}
\end{figure}

The transformation from the upper-half plane, shown in Fig. \ref{CorrugatedConformalMappingGeometry} (b), to the sloped-step geometry is given by the Schwarz-Christoffel mapping

\begin{equation}
\zeta=F(w)=K_1\int_{0}^{w}f(w')dw'+K_2,
\label{SchwarzChristoffel}
\end{equation}
where for the sloped-step geometry $f(w)=w^{\alpha_1-1}(w-1)^{\alpha_2-1}$. $K_1$ and $K_2$ are calculated from the following relations
\begin{equation}
i\delta=K_1\int_{0}^{0}f(w')dw'+K_2=K_2,
\label{eqK2}
\end{equation}
\begin{equation}
\frac{\delta}{\tan(\theta)}=K_1\int_{0}^{1}f(w')dw'+K_2.
\label{eqK1}
\end{equation}
For the repeated corrugation geometry, $f(w)$ is given by
\begin{equation} 
\label{Corrugatedf(w)}
\begin{split}
f(w)=w^{(\alpha_1-1)}(w-2N-1)^{(\alpha_3-1)}(w+2N+1)^{(\alpha_3-1)} \\
\times\prod_{n=1}^{N}(w-2n+1)^{(\alpha_2-1)}(w-2n)^{(\alpha_1-1)} \\
\times\prod_{n=1}^{N}(w+2n-1)^{(\alpha_2-1)}(w+2n)^{(\alpha_1-1)}.
\end{split}
\end{equation}
and the number of corrugations from the center, $N$, is chosen such that the corrugations extend laterally at least $50\xi$ from the origin. The current density scalar potential for the external magnetic field, is given by 
\begin{equation}
\phi_M(x,y)=\phi_M(w)|_{w=F^{-1}(x,y)}= \text{Re}(K_1J_0w)|_{w=F^{-1}}(x,y)
\label{eqJSP}
\end{equation}
and the current density scalar potential of the image vortex that accounts for the force expelling the vortex is given by \cite{kubo2015field,Hryhorenko2026ElectropolishingInducedDefect}
\begin{equation}
\phi_{I}(x,y)=\text{Re}\left( \frac{-i\phi_0}{2\pi\mu_0\lambda^2}{\ln(w-w^*_0)}\right)\biggr|_{w=F^{-1}(x,y)}.
\label{eqJSPAV}
\end{equation}
The Bean-Livingston barrier becomes unstable to vortex penetration when the sum of the forces on the vortex vanishes, $\mathbf{F_S} + \mathbf{F_M}=\mathbf{0}$, which leads to the following expression \cite{Lechner2025Nb3SnTopo}
\begin{equation}
B_s^*=\frac{2\xi|-\nabla\tilde{\Phi}_{I}(\mathbf{r_v})|}{\epsilon(\mathbf{r_v})}B_s.
\label{eq4}
\end{equation}
where $B_s = \phi_0/4\pi\lambda\xi$ is the superheating field of a perfectly flat surface in the London model, $\epsilon(\mathbf{r})=|\mathbf{J_M(r)}|/J_0=|-\nabla\Phi_{M}(\mathbf{r_v})|/J_0$ is the local current density enhancement factor, and $\mathbf{r_v}$ is the vortex nucleation position using the coherence length cutoff $\mathbf{r_v}=(\delta+\xi)\mathbf{\hat{y}}$. From Eq.~\eqref{eq4} the superheating field suppression (SFS) factor can be defined
\begin{equation}
\eta(\xi,\delta,\theta)=\frac{2\xi|-\nabla\tilde{\Phi}_{I}(\mathbf{r_v})|}{\epsilon(\mathbf{r_v})},
\label{eq5}
\end{equation}
which relates the geometrically modified superheating field to the superheating field of a flat surface. The effect of corrugated surface roughness on superheating field suppression for superconductors representing clean limit Nb, alloyed Nb and next generation materials with $\xi = 40, 13.3$ and $3$ nm is shown in Fig. \ref{EffectOfGeometryOnSFS} (a). As can be seen in Fig. \ref{EffectOfGeometryOnSFS} (a) a reduced $\xi$ reduces the superheating field suppression factor for the same defect geometry. In the limit of $N\rightarrow0$ Eq.~\eqref{Corrugatedf(w)} reduces to the single groove geometry considered by Kubo \cite{kubo2015field}. The expression for $\eta$ in Ref. \cite{kubo2015field} is an excellent approximation in the prescribed limits, as shown in Fig. \ref{ComparisonWithKubo}, but deviates from the numerical solution as the slope angle grows and $\delta/\xi$ is small.
In Fig. \ref{EffectOfGeometryOnSFS} (b) the effect of corrugation geometry on superheating field suppression in the isolated groove limit, calculated by the full numerical solution, is shown. As can be seen in Fig. \ref{EffectOfGeometryOnSFS} (b) the isolated groove geometry represents a more conservative estimate for superheating field suppression and should be used as the conservative metric for the effect of surface roughness on $B_{sh}$. The London model itself provides a conservative estimate because it neglects current-induced suppression of the superconducting order parameter. Because this model is a conservative estimate, it sets a clear target for the acceptable level of nanoscale roughness required to rule out surface roughness as a field‑limiting factor.

\begin{figure}[!h]
    \centering
    \includegraphics[width=1\linewidth]{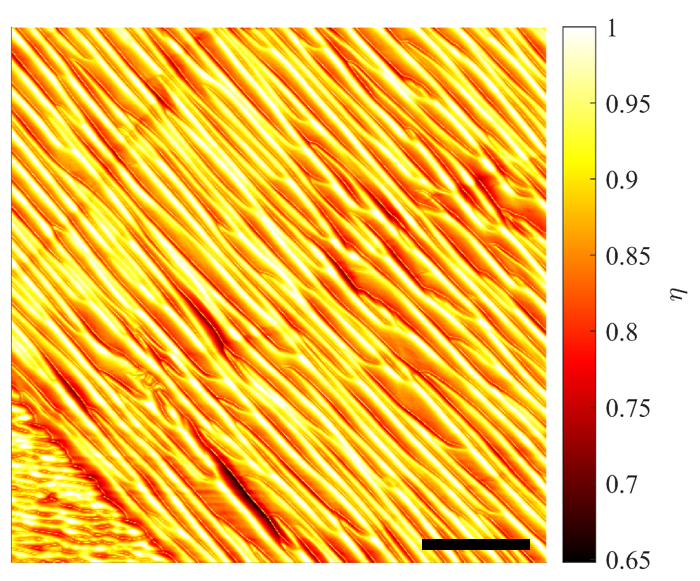}
    \caption{A representative image of the local superheating field suppression map using the conservative $N \rightarrow0$ isolated groove model. The scale bar in the image is 1 $\mu$m.}
    \label{SFSExample}
\end{figure}

An estimate for the impact of measured surface roughness on superheating field suppression can be determined by considering the local groove depth, $\delta(\mathbf{r})$ and slope angle $\theta(\mathbf{r})$ since $\eta(\delta,\xi,\theta)$. The local depth is determined by the difference between the topography and the plane that conforms to the peaks of the corrugations. This ``peak plane'' is defined at each pixel by considering the surrounding points within a radius of 160 nm, comparable to the characteristic defect wavelength observed in this work and selecting the average of the top  5 \% of z-values in that region. A superheating field suppression map derived from the topography in \ref{fig1} is shown in Fig. \ref{SFSExample}. This map was calculated using a clean limit coherence length of $\approx$40 nm \cite{McFadden2026NbCoherenceLengthAndPenetrationDepth} and reveals a pronounced suppression ($\eta \approx$0.65) for a surface whose roughness parameters are only $R_a =$ 6.7 nm and $R_q =$ 8.3 nm. This substantial reduction of the superheating field can account for the primary degradation in achievable field in Nb films. Beyond this, the inter-trough distance is very close. This close spacing makes it easy for vortices to nucleate, heat up the surrounding area, and reduce the nearby superheating field, nucleating more vortices and so on.

\subsection{\label{MFEAtCor}Magnetic Field Enhancement at Corrugations}

The size of the corrugation defects is comparable to the nanoscale defects introduced via electropolishing of Nb which cause considerable enhancement of the magnetic field \cite{Hryhorenko2026ElectropolishingInducedDefect}. To calculate the effect of these corrugation defects, the Maxwell-London equations are solved to determine the maximum local magnetic field enhancement along the surface of the superconductor. The observed corrugations are modeled by an infinitely long two-dimensional triangle-wave corrugation aligned along $\mathbf{\hat{z}}$ and an external magnetic field aligned along $\mathbf{\hat{x}}$. In this geometry only the z-component of the magnetic vector potential, $\mathbf{A}$, is relevant and the Maxwell-London equations for the magnetic vector potential become
\begin{equation}
\nabla_{xy}^2 {A_z} = 0
\label{MagneticVectorPotentialAbove}
\end{equation}
\begin{equation}
\nabla_{xy}^2 {A_z} = \frac{1}{\lambda^2}{A_z}
\label{MagneticVectorPotentialWithin}
\end{equation}
in vacuum (Eq.~\eqref{MagneticVectorPotentialAbove}) and within the superconductor (Eq.~\eqref{MagneticVectorPotentialWithin}). Since $\mathbf{A}=A_z(x,y)$, $A_z$ satisfies the Coulomb gauge. The magnetic field at the surface was calculated via $ \mathbf{B}=\nabla \times \mathbf{A}$. We define the magnetic field enhancement factor as $\beta(\mathbf{r}) = |\mathbf{B(\mathbf{r}})|/B_0$ The magnetic vector potential is continuous across the superconductor-vacuum interface. Deep ($y=-5\lambda$ for this simulation) in the superconductor $\mathbf{A}=0$. Far from the surface $\mathbf{A}=B_0y\mathbf{\hat{z}}$. At the boundary, $x=x_{min}$ and $x=x_{max}$ in and out of the superconductor, $\nabla A_z \cdot \mathbf{\hat{x}}=0$. Within this model, magnetic field enhancement depends on the relative defect size ($\delta/\lambda$) and the slope angle \cite{Hryhorenko2026ElectropolishingInducedDefect}.

\begin{figure}[!h]
\includegraphics[width=8.5cm]{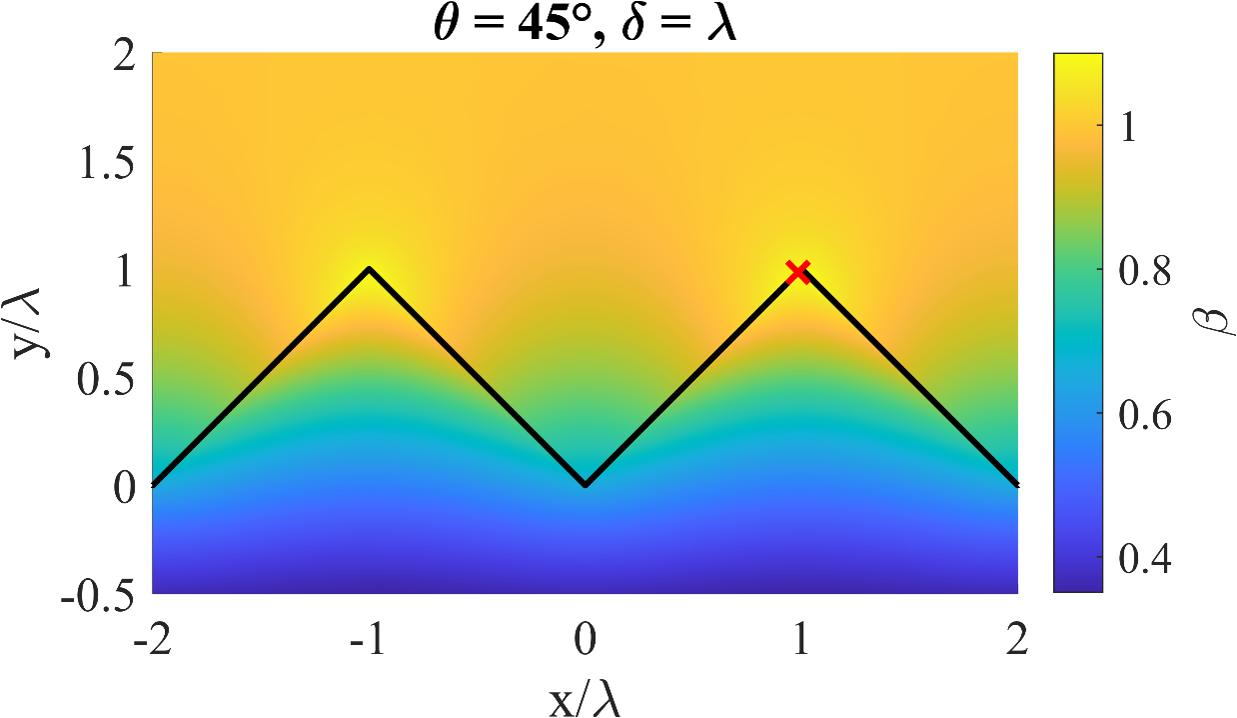}
\caption{Magnetic field enhancement due to triangle-wave corrugated surface roughness. The red cross indicates the position of the maximum magnetic field enhancement which is close to, but not exactly located at, the apex.}
\label{MFEExample}
\end{figure}

\begin{figure}[!h]
\includegraphics[width=8.5cm]{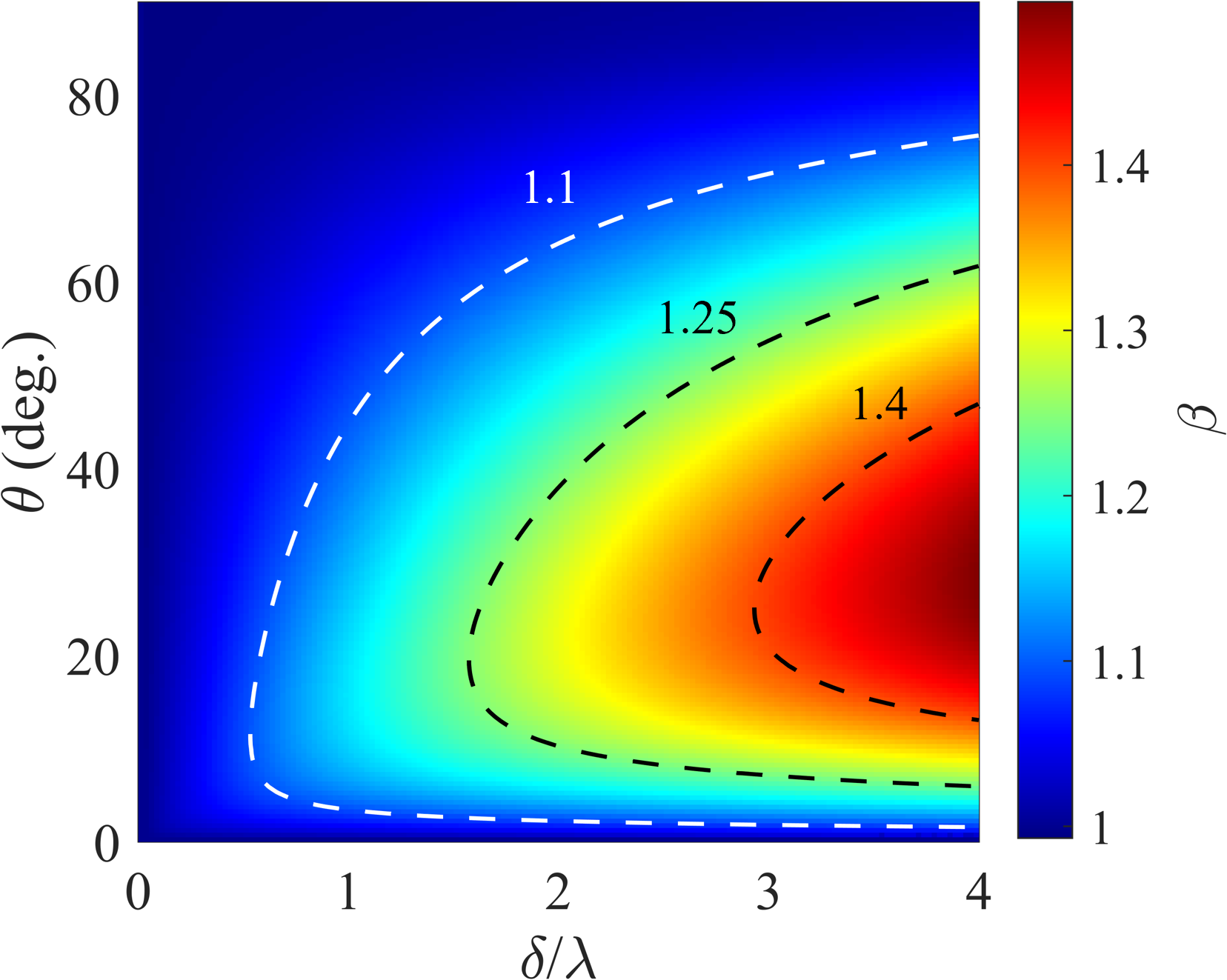}
\caption{The effect of corrugation geometry on the maximum magnetic field enhancement factor.}
\label{MFELandscape}
\end{figure}

An example of the effect of the corrugation geometry on magnetic field enhancement throughout the system is shown in Fig. \ref{MFEExample}. For the corrugated geometry, the magnetic field at the surface reaches its maximum near the geometric peak of the corrugation, as indicated by the red cross in the calculated field distribution Fig. \ref{MFEExample}. In the London model, the severity of nanoscale MFE is dictated by the relative defect size which is shown in Fig. \ref{MFELandscape}. For high slope angles, the protrusions are poorly shielded and the magnetic field enhancement decreases. For a fixed geometric defect, impurity alloying can reduce the nanoscale magnetic field enhancement by increasing the surface penetration depth. 

To make an estimate of the local field enhancement from measured topographies, a similar procedure is followed as outlined in the Sec. \ref{subsec:SFS}. Here $\beta(\theta,\delta,\lambda)$ is calculated using a clean-limit penetration depth, a groove height $\delta(\mathbf{r})$ and the local slope angle $\theta(\mathbf{r})$. To estimate the local groove height, a "trough plane" is defined at each pixel by considering the surrounding points within a radius of 160 nm, comparable to the characteristic defect wavelength observed in this work and selecting the average of the bottom  5 \% of z-values in that region. For the example topography, the London model yields a maximum MFE factor $\beta$ $\approx$ 1.2. 

\begin{figure}[!h]
\includegraphics[width=8.5cm]{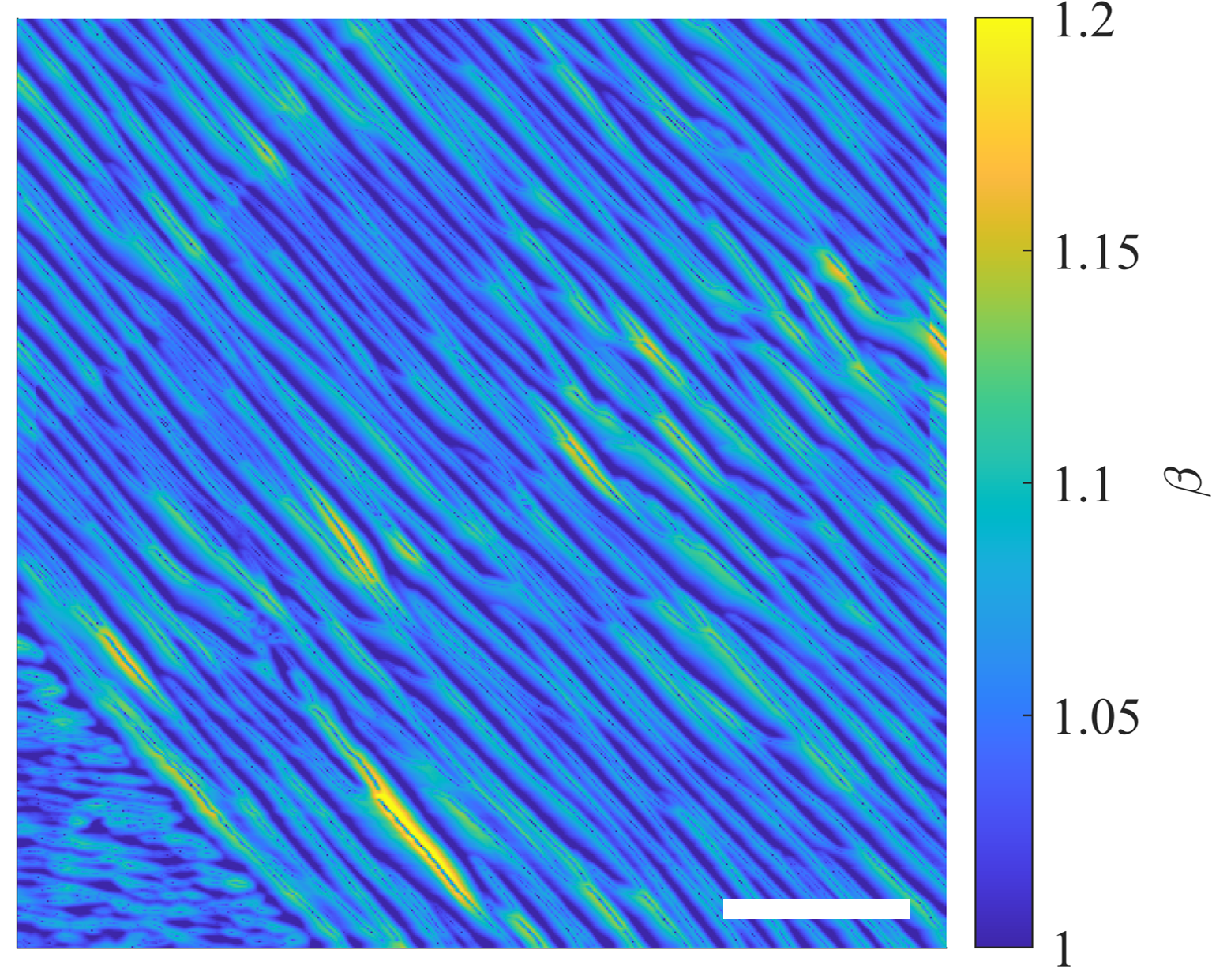}
\caption{Local magnetic field enhancement estimate for the corrugated topography in Fig. \ref{fig1} due to corrugated topography in the London model. The scale bar in the image is 1 $\mu$m.}
\label{RepresentativeMFE}
\end{figure}

\subsection{Impurity Migration at Corrugations}

The corrugation geometry hosts regions with high- and low-internal-angle which serves to locally increase and decrease impurity content near these regions. To simulate this effect we utilize Fick's second law of diffusion,

\begin{equation}
\frac{\partial c}{\partial t}  = D\nabla_{xy}^2c,
\label{diffusionEquation}
\end{equation}
where $c$ is the concentration of the impurities, and $D$ is the diffusion coefficient. Eq.~\eqref{diffusionEquation}, made dimensionless with the following transformations $\mathbf{r}/L_D \rightarrow \Tilde{\mathbf{r}}$, $L_D\nabla \rightarrow \Tilde{\nabla}$, $t/t_D \rightarrow \Tilde{t}$ and $c/c_0 \rightarrow \Tilde{{c}}$. Here $c_0$ is defined as the concentration at the surface after 1$t_D$ on a flat surface. The diffusion of impurities in this model depends on the slope angle and the normalized corrugation depth, $\delta/L_D$. The diffusion length is defined by $L_D=\sqrt{Dt_D}$ which also defines $t_D$.

\begin{figure}[!h]
\includegraphics[width=8.5cm]{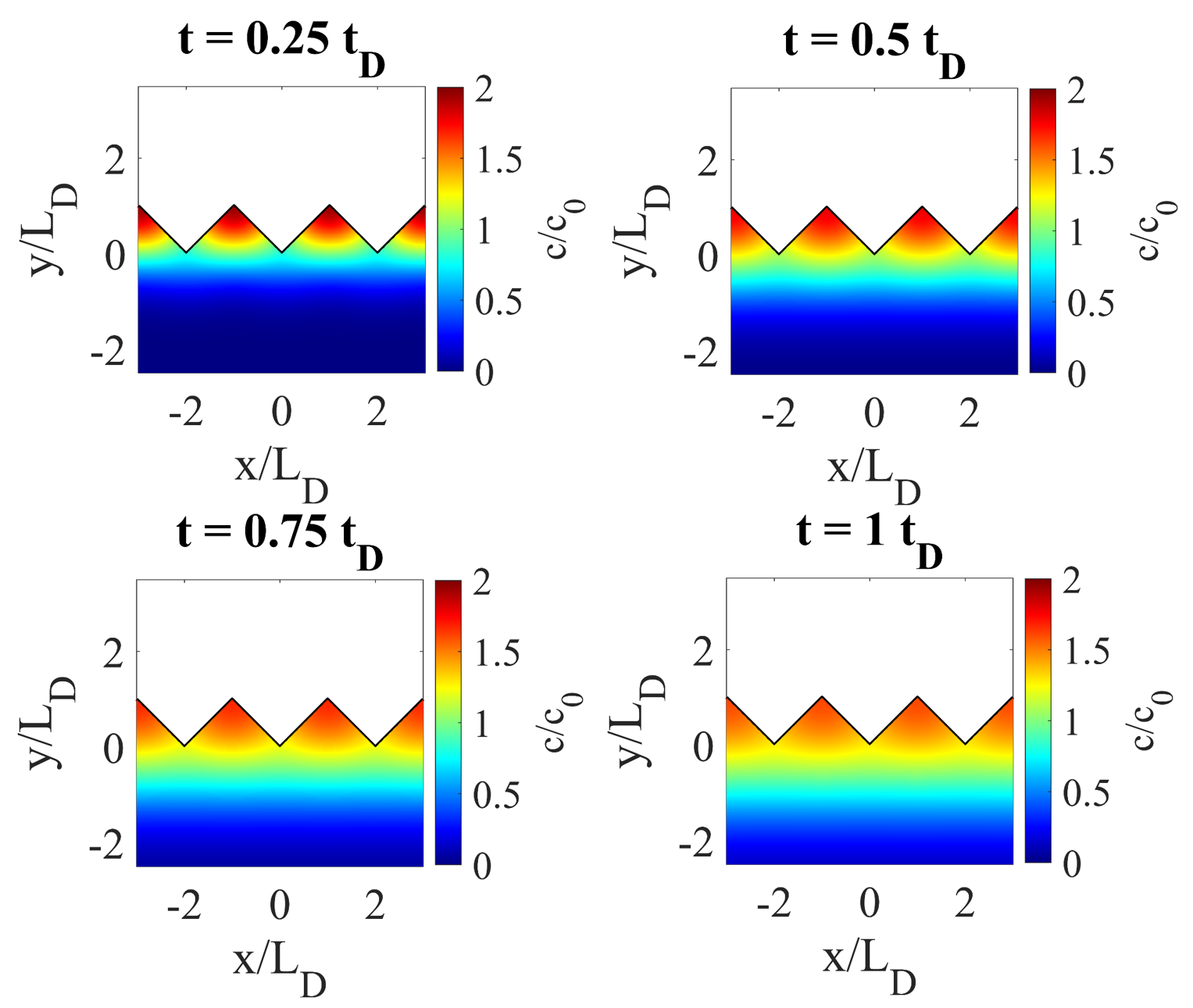}
\caption{Snapshots of impurity diffusion around corrugations in the finite source model. The four panels display the normalized impurity concentration $c/c_0$ at progressive times $t=0.25t_D$, $t=0.5t_D$, $t=0.75t_D$, and $t=1t_D$, where $t=t_D$ is the characteristic diffusion time.}
\label{DiffusionSnapshots}
\end{figure}

Shallow impurity profiles have demonstrated high accelerating fields in either low-temperature/mid-T baking or nitrogen infusion \cite{Ciovati2004EffectOfLTbaking,Bafia2021HighGradientsRoleofOxygen,Khanal2023InsightLTBDuration,grassellino2017unprecedented,Dhakal2018EffectOfLTBinN,Ito2021FurnaceBaking}. Depending on the method of impurity introduction the impurity flux through the surface can take a wide variety of forms. In this work, three simple scenarios for the shallow migration of impurities on a length scale comparable to the corrugation depth are considered: a finite impurity source, an infinite source, and constant flux of impurities. A detailed description of the boundary conditions for these scenarios can be found in the supplemental material. The results of the finite impurity source model \cite{Ciovati2006ImprovedODiffusion, lechner2021rf} are presented here. The results of the other scenarios can be found in the supplemental material. Snapshots of the impurity diffusion process near the topographic defect are presented in Fig. \ref{DiffusionSnapshots}. It can be seen that the effective impurity concentration is diminished near bottom of the sloped corrugation as a consequence of the greater volume for the impurities expand into. Conversely, the top of the topographic defect can host an enhanced impurity concentration compared to the regions far from the defect.

\begin{figure}[!h]
\includegraphics[width=8.5cm]{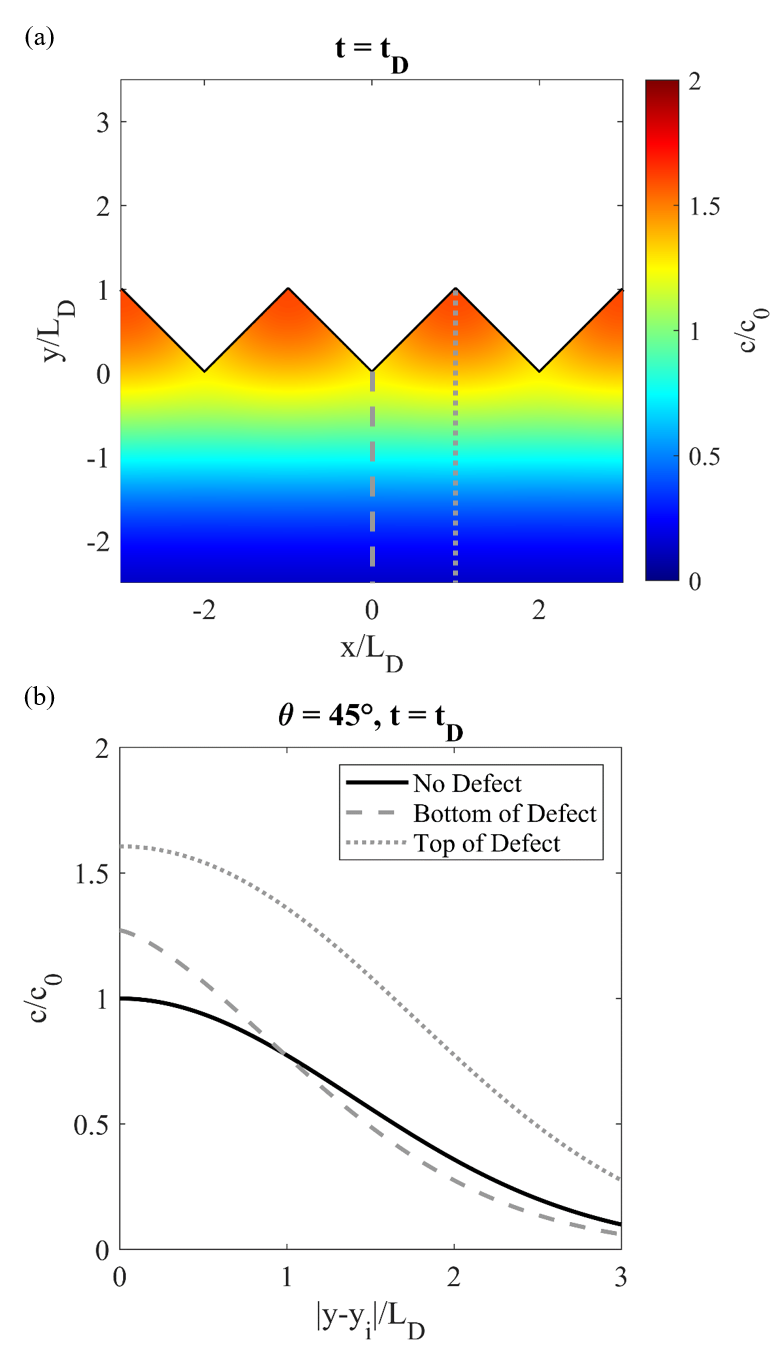}
\caption{(a) Solution of Eq.~\eqref{diffusionEquation} the dashed and dotted lines extending toward the bulk indicates the location of the line profile plotted in (b). (b) Comparison between impurity profiles at the top and bottom of the defect as well as far from the defect.}
\label{FiniteSourceExample}
\end{figure}

\begin{figure}[!h]
\includegraphics[width=8.5cm]{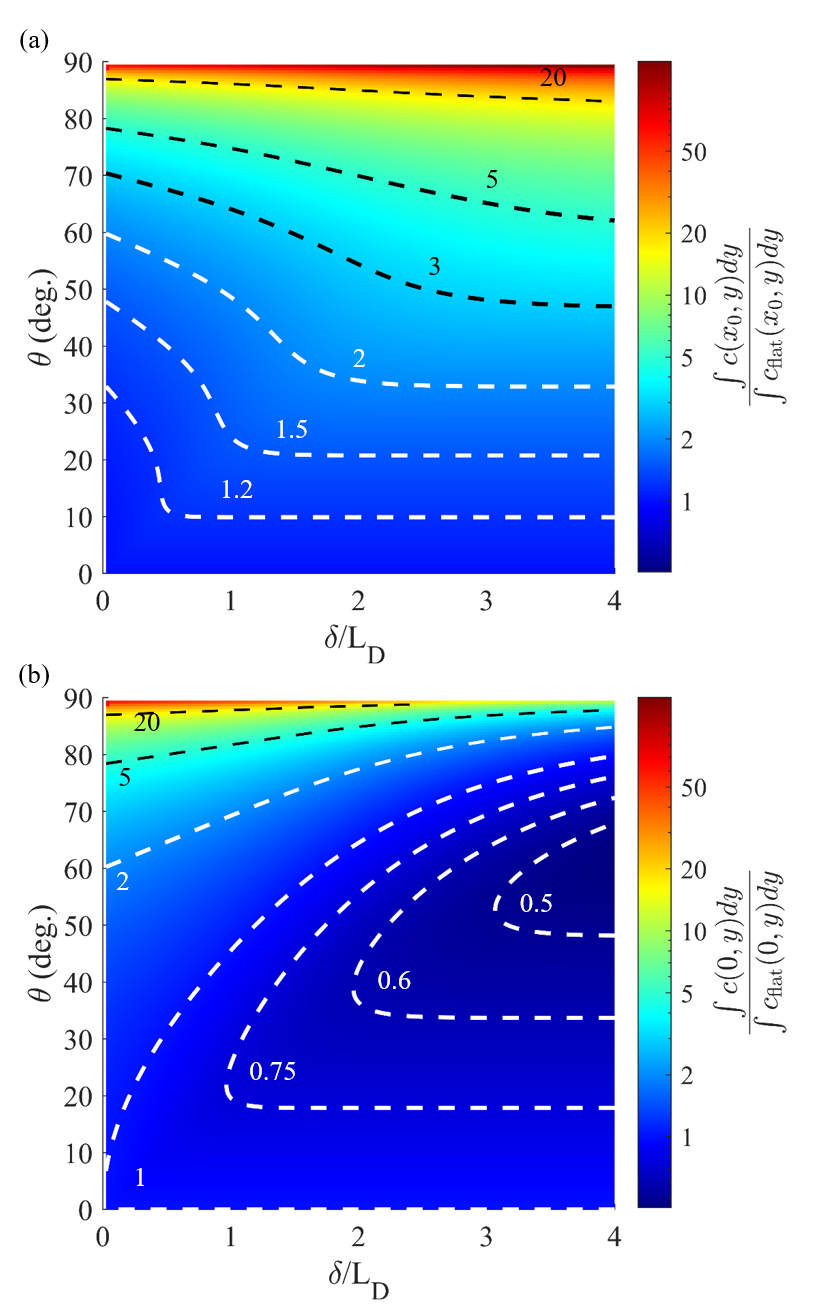}
\caption{The effect of topography on relative impurity dose at $t/t_D=1$ at (a) the corrugation top and (b) corrugation trough.}
\label{FiniteSourceLandscape}
\end{figure}

An example of the effect of nanoscale corrugated roughness on impurity diffusion for a $\theta=45$°, $\delta=L_D$ at $t=t_D$ is shown in Fig. \ref{FiniteSourceExample} (a). The surface geometrically confines the impurity content and enhances the impurity concentration near the peaks. In this example even the impurity content at the groove bottoms can be enhanced, as shown in Fig. \ref{FiniteSourceExample} (b), which occurs as a result of the increase in surface area. The effect of step geometry on the relative impurity dose, $\int c(x,y)\,dy/\int c_{flat}(0,y)\,dy$, is shown in Fig. \ref{FiniteSourceLandscape} (a) and (b) for the top of the corrugation and the bottom. At the corrugation top, the impurity content is always enhanced due to the reduced internal angle, while for the trough, at $t_D$, the impurity content can be either increased or decreased. For nanoscale high-slope angle serrations the relative surface area $A_{r}/A_{s}=\sec(\theta)$ where $A_{r}$ is the area of the rough surface and $A_{s}$ is the area of the smooth surface and impurity content is strongly increased. The effective impurity dose is increased everywhere when the groove's relative dose exceeds unit indicated by the contour line in Fig. \ref{FiniteSourceLandscape} (b). It is interesting to note the variety of impurity doses possible in the corrugated roughness geometry since impurity content may strongly affect vortex nucleation in the superconductor-superconductor bilayer theory \cite{kubo2015field,KuboMultilayerReview2016} or hydride precipitation \cite{Pfeiffer1976TrappingHydrogen}. In some geometries, the total impurity dose can exceed the flat smooth surface and may enhance the effectiveness of both mechanisms.

\subsection{Mitigation of MFE via Shallow Impurity Diffusion}

Shallow impurity profiles are particularly relevant to the high-field performance of Nb SRF cavities. High-field $Q$ slope, a phenomenon characterized by a rapid decrease in $Q_0$ above accelerating gradients between approximately 20 and 30 MV/m \cite{Ciovati2004EffectOfLTbaking}, can be suppressed by low-temperature baking at temperatures around $100$ - $160$ °C \cite{bauer2006QSlopeCollection}. The effectiveness of this treatment has been associated with a reduction in the near surface electron mean free path and the redistribution of O over a depth comparable to the RF screening length. Nitrogen infusion \cite{grassellino2017unprecedented} similarly introduces a shallow impurity profile \cite{koufalis2019EffectofLowTemperatureInfusionHeatTreatmentsand20NDoping} and has produced high accelerating fields with suppressed high-field $Q$ slope. Although the impurity species and processing conditions differ, both treatments produce spatial variations in the mean free path and London penetration depth near the surface. Previous work showed that such profiles can redistribute the Meissner screening current and reduce its peak value, thereby helping to preserve the Bean-Livingston barrier \cite{Lechner2024OxideDisScenarios}. In that work, the corresponding increase in the attainable field was estimated by scaling it inversely with the calculated peak screening current density. Calculations that treat vortex penetration more directly predict a substantially smaller increase in the superheating field of $\approx 3-9\%$ \cite{Pathirana2023SuperheatingFieldNanostructured,Wallace2026FirstVortex} than the simplistic estimate from supercurrent suppression alone \cite{Lechner2024OxideDisScenarios}. Here, we consider a complementary consequence of the same modification to the screening response: the effect of a shallow impurity profile on nanoscale magnetic field enhancement at a corrugated surface.

Increasing the penetration depth relative to defect size reduces the effective magnetic field enhancement, as shown in Fig. \ref{MFELandscape}. The penetration depth need not be uniform to achieve the same effect. The effect of shallow impurity diffusion on magnetic field enhancement for a spatially varying London penetration depth is determined by solving 
\begin{equation}
\nabla_{xy}^2 {A_z} = \frac{1}{\lambda^2(x,y)}{A_z}.
\label{MagneticVectorPotentialWithinLambdaSpatiallyVarying}
\end{equation}
inside the superconductor. For the two-dimensional geometry considered here, Eq.~\eqref{MagneticVectorPotentialWithinLambdaSpatiallyVarying} follows directly from the generalized London equation for an inhomogeneous superconductor \cite{Kogan2011MeissnerInhomogeneousPenetrationDepths}. Recent comparisons with microscopic Eilenberger calculations support the generalized London treatment underlying Eq.~\eqref{MagneticVectorPotentialWithinLambdaSpatiallyVarying} for depth-dependent nonmagnetic impurity profiles, finding near equivalent magnetic field and screening current distributions at fields sufficiently below the superheating field \cite{herrero2026effectnonhomogeneousnonmagneticimpurities}. However, near the superheating field, the microscopic calculations deviate from the generalized London model because they capture the nonlinear Meissner response. The vacuum equation and boundary conditions remain as described in Sec. \ref{MFEAtCor}.

To simulate this effect, a finite source oxygen impurity diffusion scenario is considered as described by Ciovati \cite{Ciovati2006ImprovedODiffusion} and using oxygen diffusion parameters as measured in Ref. \cite{lechner2021rf}. For simplicity, we consider only an initial interstitial O dose, $v_0$, in the metal as the main contributor to shallow impurity diffusion. To connect the O distribution to the local screening response, the electron mean free path was estimated from the impurity induced change in resistivity following Ref. \cite{Lechner2024OxideDisScenarios},
\begin{equation}
\ell(x,y)=\frac{\sigma}{a c(x,y)},
\label{OxygenMeanFreePath}
\end{equation}
where $\sigma=0.37\times10^{-15}$ $\Omega$ m$^2$ \cite{Goodman1968InfluenceOfExtendedDefectsOnSCInNb} and $a c$ is the increase in resistivity produced by interstitial O, $a=4.5\times10^{-8}$ $\Omega$ m \cite{Schulze1981PrepOfUltraHighPurityNb}, and $c(x,y)$ is the local O concentration in at. \%. The local London penetration depth was then approximated as \cite{Tinkham1958PenetrationDepthInAlloys}
\begin{equation}
\lambda(x,y)=\lambda_0\sqrt{1+\frac{\xi_0}{\ell(x,y)}},
\label{OxygenPenetrationDepth}
\end{equation}
where $\lambda_0=30$ nm and $\xi_0=40$ nm are the clean-limit London penetration depth and coherence length used throughout this work. A uniform background O concentration of 10 ppma was included in $c(x,y)$. 

\begin{figure}[!h]
\includegraphics[width=8.5cm]{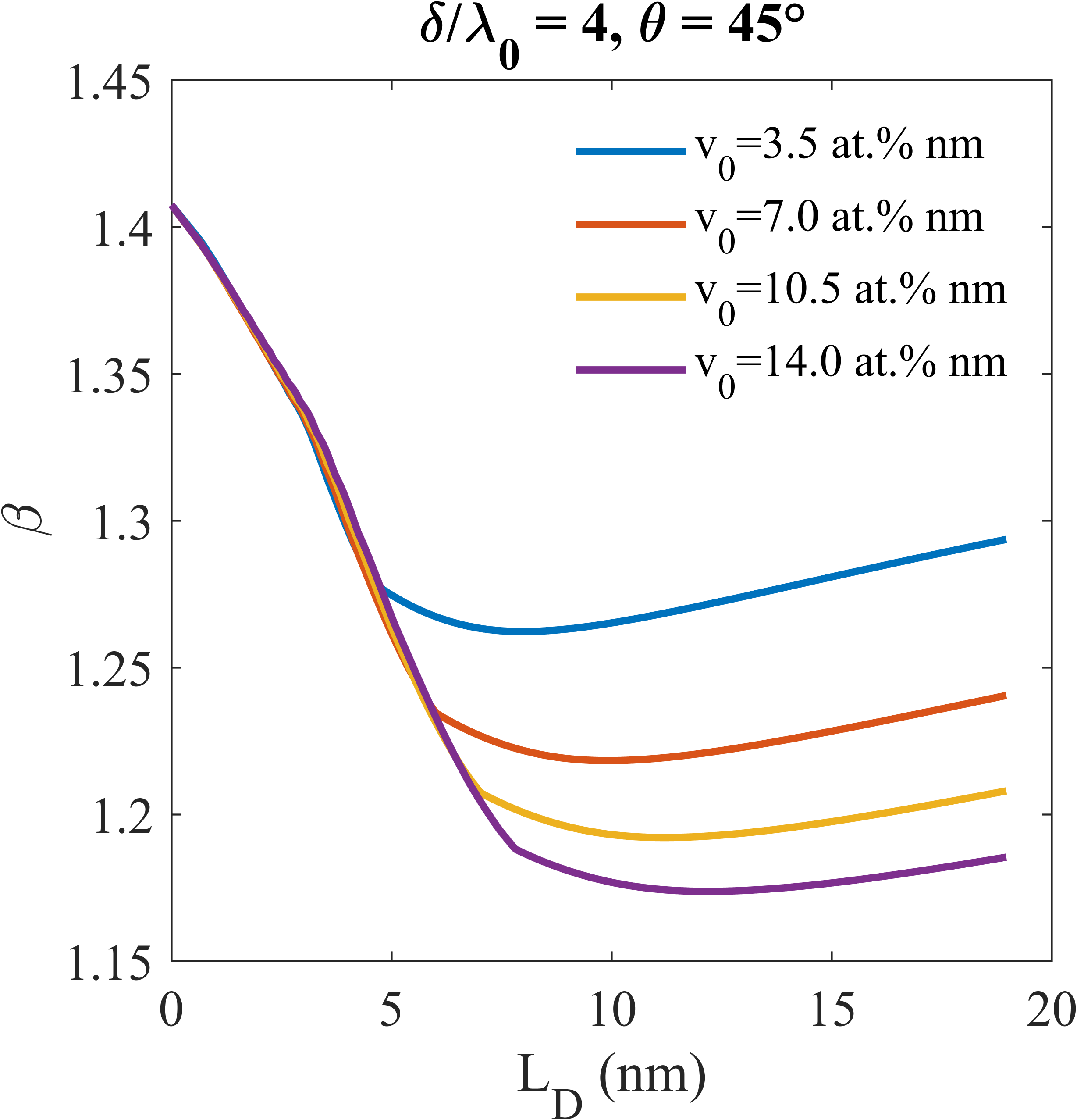}
\caption{The effect of shallow impurity migration on MFE for $\delta/\lambda_0=4$ and $\theta=45$° at various impurity doses.}
\label{MFEvsDiffusionLength}
\end{figure}

Fig. \ref{MFEvsDiffusionLength} shows the effect of shallow O diffusion on MFE for different O diffusion lengths for multiples of the initial impurity dose $v_0$. As impurities migrate, the lengthening screening length reduces the peak magnetic field enhancement, resulting in a minimum in MFE. Upon increasing the diffusion length further MFE slowly increases again due to the diminishing impurity content near the surface. The significance of this result is that shallow impurity profiles can substantially modify nanoscale magnetic field enhancement. An argument previously advanced against the magnetic field enhancement model of the high-field Q slope, and its ability to account for changes following low temperature baking, is that baking does not alter the surface geometry and therefore cannot alter $\beta$ \cite{bauer2006QSlopeCollection}. However, this reasoning is valid only when all locally relevant geometric length scales are much larger than the London penetration depth. It breaks down when dimensions such as groove depth, spacing, or local radius of curvature are comparable to $\lambda$. Since buffered chemical polished cavities have feature sizes that are much greater than $\lambda$, this scale dependence likely explains the differing responses of buffered chemical polished and electropolished bulk Nb cavities to low-temperature baking. In this regime, $\beta$ depends on normalized defect dimensions such as $\delta/\lambda$. Consequently, an impurity-induced increase in $\lambda$ reduces the effective defect size and can suppress magnetic field enhancement without any change in surface roughness. Similar reductions are obtained for shallow impurity profiles applied to the characteristic sloped-step defects of electropolished Nb surfaces; a detailed analysis of these results will be reported elsewhere.

For the finite source model with an initial oxygen dose of $v_0=3.5$ at. \% nm, diffusion around a defect with $\delta=4\lambda_0=$120 nm and $\theta=$45° reduces $\beta$ from 1.41 to a minimum of approximately 1.26, where $\lambda_0$ is the clean-limit London penetration depth. Increasing the initial dose to $4v_0$ further reduces the minimum $\beta$ to approximately 1.17, demonstrating the importance of the oxygen loading at the surface. To illustrate the potential effect on cavity performance, consider a limiting local magnetic field of approximately 210 mT, corresponding to 50 MV/m in a TESLA-shaped cavity \cite{AuneSuperconducting2000}. If magnetic field enhancement alone determines the attainable gradient, $\beta=1.41$ reduces this gradient to 35.5 MV/m, whereas reducing $\beta$ to 1.26 or 1.17 increases it to 39.7 or 42.7 MV/m, respectively. Determining the initial near-surface oxygen dose is therefore important for quantifying this effect, particularly because doses as high as 13 at. \% nm have been reported \cite{Lechner2024OxideDisScenarios}. This reduction in $\beta$ will be further enhanced when considering the nonlinear Meissner effect. Furthermore, shallow impurity profiles can provide a secondary enhancement through enhancement of the superheating field which is estimated to be $\approx3-9\%$ \cite{Pathirana2023SuperheatingFieldNanostructured,Wallace2026FirstVortex}. Depending on the impurity loading, these combined effects may represent the primary mechanism by which low temperature baking and nitrogen infusion improve high-field SRF cavity performance.
 
\begin{figure}[!h]
\includegraphics[width=8.5cm]{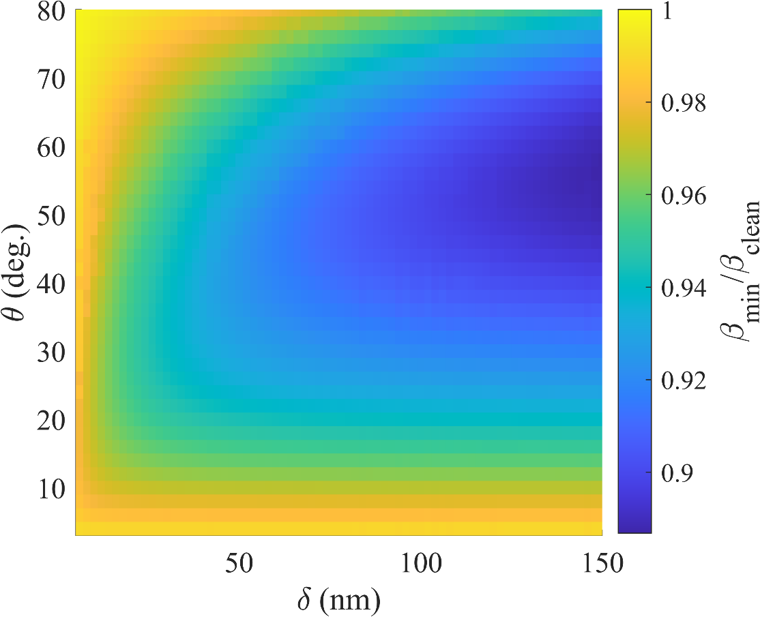}
\caption{Effect of defect geometry on the minimum normalized magnetic field enhancement, $\beta_{\min}/\beta_{\mathrm{clean}}$, for $v_0 =$ 3.5 at. \% nm.}
\label{MFEDiffusionLengthLandscape}
\end{figure}

The effect of defect geometry on the minimized magnetic field enhancement for an impurity dose of $v_0=3.5$ at. \% nm is shown in Fig. \ref{MFEDiffusionLengthLandscape}. Here, $\beta_{\min}$ is the minimum value attained over the diffusion lengths considered, and $\beta_{\mathrm{clean}}$ is the enhancement factor before impurity migration for the same defect geometry. The mitigation is geometry dependent, with $\beta_{\min}/\beta_{\mathrm{clean}}$ reaching approximately 0.89 for deeper corrugations with intermediate slope angles, while remaining close to unity for shallow or low slope angle defects.

\section{Discussion}
The central result of this work is that the performance-relevant roughness of Nb films on Cu is not adequately described by conventional roughness metrics. The measured surfaces exhibit a corrugated morphology with characteristic amplitudes and lateral length scales comparable to the characteristic superconducting length scales in Nb. This is important because vortex nucleation is not solely controlled by the amplitude of the height variation, which is quantified by metrics like $R_a$, but also controlled by the local slope angle, corrugation depth, and the ratios $\delta/\lambda$ and $\delta/\xi$. In this sense, films that appear smooth by average roughness metrics may still contain mesoscopic topographic features that are severe from perspective vortex nucleation. This distinction is made explicit by the idealized triangle-wave geometry considered here. For this geometry, simple roughness metrics, like average roughness, is independent of the slope angle for fixed corrugation depth ($R_a = \delta/4$). Therefore, two corrugated surfaces with identical $R_a$ can have substantially different local slopes and, consequently, substantially different magnetic field enhancement and vortex-entry thresholds. Our results provide a simple explanation for why roughness trends inferred from buffered chemical polished and electropolished bulk Nb cavities fail when applied to Nb/Cu films. The empirical comparison between BCP and EP correctly captures that smoother bulk Nb surfaces often perform better, but it does not establish $R_a$ or $R_q$ as a universal predictor of SRF-relevant surface quality, especially when the structure of defects is substantially different. For structured morphologies such as the corrugations studied here, local geometric quantities provide a more direct connection to field limitation than the average absolute height deviation from a mean plane.

The superheating field suppression calculations show that corrugated grooves can \emph{strongly} reduce the effective surface barrier to vortex entry. In the extreme type-II limit, the superheating suppression factor ($\eta$) depends on the corrugation depth, slope angle, and coherence length. For the same defect geometry, increasing $\delta/\xi$ and increasing slope angle more strongly suppress the geometrically modified superheating field. For the representative measured topography, applying the conservative isolated groove limit gives local values as low as ($\eta \approx $0.65). Such a reduction is substantial enough that these nanoscale corrugated topographic defects cannot be dismissed as an insignificant field limiter. 

The magnetic field enhancement calculations provide a complementary view of the same geometric problem. Solving the Maxwell-London equations for representative corrugated geometries shows that the local field can be enhanced near the corrugation peaks. For the measured topography considered here, the London calculation gives a maximum magnetic field enhancement factor of approximately ($\beta \approx $1.2). This is smaller than the SFS degradation, but still relevant for SRF operation, bringing the superconductor closer to vortex-entry or thermal-instability thresholds. 

It is important to distinguish the two models used to calculate magnetic field enhancement and superheating field suppression. The magnetic field enhancement model describes the redistribution of magnetic field caused by the rough surface in the London model. In this case, the external magnetic field points against the roughness profile. The superheating field suppression model describes the reduction of the vortex-entry field, associated with the corrugated geometry. In this case, the external magnetic field points parallel to the roughness profile grooves. The external magnetic field for each of these scenarios is perpendicular to each other. This means that these quantities should not be treated as automatically independent multiplicative penalties unless a specific geometry causes both to coincide. Grain boundary triple junctions are good examples of the type of defects that can cause coincidence \cite{Lechner2025Nb3SnTopo}. An obvious nanoscale geometric defect inducing a combined $\eta/\beta$ field reduction was not identified here. Regardless, the multiscale nature of roughness of the surface will introduce simultaneous magnetic field enhancement and superheating field suppression like that due to larger scale pits, protrusions or grain boundary steps.

Finally, a useful comparison can be made with a previous analysis of grain boundary sloped-step defects introduced during electropolished of bulk Nb \cite{Hryhorenko2026ElectropolishingInducedDefect} using the same theoretical framework. First, it is important to note that for wide and centered measurements over grain boundaries, which have step-function-like surface roughness, the surface roughness is $R_a=\delta/2$, where $\delta$ is the height of the step while corrugations have $R_a=\delta/4$. Already, when comparing corrugated defects and sloped steps, we can see for defects of equal height and slope angle that $R_a$ does not provide meaningful insight into which surface is better for high-field performance. A naive comparison of $R_a$ in this situation would indicate that the film has a better surface, yet this is far from the case. Despite having the lower average surface roughness the film's defect structure is substantially different, hosting a \emph{larger internal angle} which degrades the surface barrier much more than the sloped-step geometry. This provides insight into the counterintuitive trends observed on differently processed Nb. Isolated groove defects, which are the most detrimental, are even more difficult to interpret through $R_a$ since $R_a\rightarrow0$ as the scan size grows. Now, considering superheating field suppression, the most severe measured topographic defects produced superheating field suppression factors of approximately $\eta \approx$0.77-0.80 \cite{Hryhorenko2026ElectropolishingInducedDefect}. In contrast, the representative Nb film on Cu topography examined here contains regions with $\eta$ as low as approximately 0.65. Thus, the characteristic corrugations of the Nb film can produce substantial suppression of the surface barrier than even the most severe defects identified in the previous bulk Nb measurements. Moreover, whereas the low-$\eta$ regions in bulk Nb are directly caused by grain-boundary steps, the corrugations in the Nb/Cu film can be ubiquitous across the scan area. This may be related to the ubiquitous low-field vortex nucleation observed in recent near-field microwave microscopy measurements \cite{Wang2024NonlinearMicrowaveResponseOfSRFNbFilms}. Therefore, not only do the Nb/Cu films differ from electropolished bulk Nb in the severity of individual geometrical defects, but also in the prevalence of regions susceptible to low-field vortex penetration.

\section{Conclusion}
We further characterized the corrugated topography common among Nb films deposited on Cu. This topography hosts large slope angle, sharp grooves and corrugation depths comparable to the London penetration depth and coherence length.  Using the London model framework, we show that such corrugations produce significant local magnetic field enhancement and superheating field suppression. The maximum magnetic field enhancement factor $\beta$ reaches values of approximately 1.2 at the peaks of the corrugations, driven primarily by the interplay between the intergranular step height $\delta$  (on the order of the London penetration depth $\lambda$) and the local slope angle $\theta$. This work demonstrates that despite the apparently low average roughness metrics, the surface morphology of Nb films are not smooth by the metrics that matter for superconducting radio frequency cavities. 

These results clarify why Nb/Cu cavities exhibit performance that deviates from the intuitive surface roughness trends observed in bulk Nb cavities processed by buffered chemical polishing or electropolishing. Conventional surface roughness metrics such as average roughness $R_a$ are insufficient to capture the performance relevant topographic features, the local slope angle and corrugation geometry must be considered explicitly. The characteristic corrugations identified in Nb films on Cu therefore represent a key factor limiting the achievable accelerating gradient and quality factor in these composite structures. 

The findings underscore the critical importance of controlling not only the average surface roughness but also the detailed mesoscopic topography during Nb film growth for high-performance SRF applications. Future work should explore deposition strategies (e.g. ion energy, substrate roughness, substrate temperature, or post deposition processing) aimed at minimizing high slope angle corrugations and the groove depths. This study provides a conservative framework for assessing how as‑measured Nb/Cu surface topography influences SRF‑relevant performance limits. Additionally, extending these analyses to include realistic impurity profiles and thermal feedback effects will further bridge the gap between fundamental surface science and practical SRF cavity performance. 

Finally, geometric effects at these corrugations influence near surface impurity distributions. Finite source diffusion simulations show that the corrugated geometry leads to nonuniform impurity concentrations, with reduced impurity doses at the base of the sloped corrugations where vortex nucleation is most likely to occur. This may affect the effectiveness of hydride blocking or extending the field-limits via a dirty-superconductor/superconductor bilayer. More broadly, increasing $\lambda$ through shallow diffusion of impurities can reduce the enhancement of the magnetic field due to the nanoscale surface roughness. This result recasts impurity profile engineering as a means of mitigating the electromagnetic consequences of nanoscale surface roughness.

\section{Acknowledgments}
This material is based upon work supported by the U.S. Department of Energy, Office of Science, Office of Nuclear Physics under Contract No. 89243126CSC000213. B.F.R. is supported by Department of Energy, Applied Innovative Traineeships in Accelerators for Nuclear Physics Research, DE-SC0022537 Support for T.S.C. was provided by the U.S. National Science Foundation Research Experience for Undergraduates at Old Dominion University Grant No. 2348822.

\bibliographystyle{apsrev4-1.bst}
\bibliography{apssamp}

\end{document}